\documentclass{article}

 \usepackage[dblblindworkshop, final]{neurips_2025}
\workshoptitle{Deep Learning for Code in the Agentic Era}

\usepackage[utf8]{inputenc} 
\usepackage[T1]{fontenc}    
\usepackage{hyperref}       
\usepackage{url}            
\usepackage{booktabs}       
\usepackage{amsfonts}       
\usepackage{nicefrac}       
\usepackage{microtype}      
\usepackage{xcolor}         

\usepackage{soul}
\usepackage{xspace}
\usepackage{graphicx}
\usepackage{listings,multicol}
\usepackage[caption=false]{subfig}  
\usepackage[export]{adjustbox}
\usepackage{amsmath}
\usepackage{amssymb}
\usepackage{textcomp}
\usepackage{listings}
\usepackage{algorithm}
\usepackage[noend]{algpseudocode}
\usepackage{hyperref}
\usepackage{cleveref}
\usepackage{booktabs}
\usepackage{multirow}
\usepackage{enumerate}
\usepackage[shortlabels]{enumitem}
\usepackage{makecell}
\usepackage{spverbatim}
\usepackage{fancyvrb}
\usepackage{tabularx}
\usepackage{tablefootnote}
\usepackage{tcolorbox}
\usepackage{color}
\usepackage{colortbl}
\usepackage{comment}

\newcommand{\ie}{\emph{i.e.,}\xspace}
\newcommand{\eg}{\emph{e.g.,}\xspace}

\newcommand{\dsreal}{Diff-XYZ\xspace}

\setcitestyle{square,numbers}

\title{\dsreal: A Benchmark for Evaluating Diff Understanding}

\author{%
  Evgeniy Glukhov \\
  JetBrains Research\\
  Amsterdam, the Netherlands \\
  \texttt{evgeniy.glukhov@jetbrains.com} \\
  \And
  Michele Conti \\
  JetBrains Research \\
  Amsterdam, the Netherlands \\
  \texttt{michele.conti@jetbrains.com} \\
  \And
  Egor Bogomolov \\
  JetBrains Research \\
  Amsterdam, the Netherlands \\
  \texttt{egor.bogomolov@jetbrains.com} \\
  \And
  Yaroslav Golubev \\
  JetBrains Research \\
  Belgrade, Serbia \\
  \texttt{yaroslav.golubev@jetbrains.com} \\
  \And
  Alexander Bezzubov \\
  JetBrains Research \\
  Amsterdam, the Netherlands \\
  \texttt{alexander.bezzubov@jetbrains.com}
}

\begin{document}

\maketitle

\begin{abstract}
Reliable handling of code diffs is central to agents that edit and refactor repositories at scale. 
We introduce \emph{\dsreal}, a compact benchmark for code–diff understanding with three supervised tasks:
\textit{apply} (old code + diff $\rightarrow$ new code), 
\textit{anti-apply} (new code -- diff $\rightarrow$ old code), and 
\textit{diff generation} (new code -- old code $\rightarrow$ diff). 
Instances in the benchmark are triples $\langle \textit{old code}, \textit{new code}, \textit{diff} \rangle$ drawn from real commits in CommitPackFT, paired with automatic metrics and a clear evaluation protocol.
We use the benchmark to do a focused empirical study of the unified diff format and run a cross-format comparison of different diff representations.
Our findings reveal that different formats should be used depending on the use case and model size. 
For example, representing diffs in search-replace format performs best for larger models across most tasks, while structured udiff variants offer similar but slightly weaker performance. In contrast, smaller open models benefit little from any formatting choice.
The \dsreal benchmark is a reusable foundation for assessing and improving diff handling in LLMs that can aid future development of diff formats and models editing code.
The dataset is published on HuggingFace Hub: \url{https://huggingface.co/datasets/JetBrains-Research/diff-xyz}.

\end{abstract}

\section{Introduction}
Modern code-capable language models increasingly interact with repositories by both analyzing and generating code diffs.
In tasks such as issue resolution \citep{swe-rebench, swebench}, CI build repair \citep{LCA}, or bug repair \citep{defects4j}, the model's task is to produce a patch for an existing codebase that is then evaluated for correctness.
Tasks like commit message generation treat diffs as an input that a model should analyze to generate a summary \citep{CMG, CMG-2}.

There exists a multitude of formats for representing code diffs.
Classical diff formats include the
(i) normal format -- changes without surrounding context,
(ii) context format -- changes with surrounding lines, and
(iii) unified format\footnote{Throughout the work, we use \emph{unified diff} or \emph{udiff} for all the diffs that use the unified diff syntax.} -- a compact, context-carrying variant with \texttt{@@} hunk headers and \texttt{+}/\texttt{-} prefixes (see the example in Figure~\ref{fig:udiff-example}).
These three formats are standardized in GNU diffutils \citep{diffutils} and remain the basis for machine-applicable patches.
In modern tooling, \texttt{git diff} produces a Git patch similar to unified diff but with additional Git-specific headers.

While benchmarks such as SWE-bench evaluate patches in the unified diff format, the choice of the diff representation used by an LLM can influence the quality of generations and their cost.
For example, Aider reports that switching from a search/replace scheme to unified diffs cuts unhelpful outputs and raises benchmark scores \citep{Aider-udiff}.
OpenAI trains GPT-4.1 to generate and apply diffs in a recommended \textit{V4A} patch format, and reports large gains on Aider’s polyglot diff benchmark \citep{aider-polyglot, gpt4.1}.
Existing works \citep{refactai, demystifying, swe-agent} report using different approaches, such as working with unified diffs, performing rewrites on specified ranges, supporting search-replace formats that specify anchored replacements.

Despite this centrality, current evaluations make it hard to isolate the effect of diff format from other factors like retrieval or tool use.
\citet{wang2025solved} analyze SWE-bench and highlight that there exist distinct failure modes: a patch may fail because it is syntactically malformed, because it does not apply to the intended context, or because it applies but does not fix the issue. 
This shows that patch outcomes depend on more than just representation, making it difficult to isolate and evaluate the impact of diff formats within such end-to-end benchmarks.

To enable studies targeting diff representations, we introduce \dsreal, a focused lightweight benchmark for exploring how well models work with various diff formats. 
Our benchmark allows varying the diff representation while keeping the rest of the context fixed.
\dsreal consists of three synthetic tasks that represent three possible problems of finding an unknown in the equation $\textit{diff} = \textit{new code} - \textit{old code}$. 
We hypothesize that solving complex generation and analysis tasks involving code diffs requires first mastering simpler variants of the problem. These include cases where two elements of the equation $\textit{new code} - \textit{old code} = \textit{diff}$ are provided and the third must be inferred.
Rewriting it as $X - Y = Z$, we obtain the following list of tasks.

\begin{enumerate}[label*={},label=\Alph*.]
    \item[X.] \textit{Apply Task} --- \textit{new code} is unknown. This tests format obedience and character-level fidelity. The model must analyze a diff in a given format and realize the exact edits, including whitespace and ordering.
    \item[Y.] \textit{Anti-Apply Task} --- \textit{old code} is unknown. This probes invertibility and losslessness of the chosen format as processed by the model. The model must reconstruct deletions and replacements precisely and align modified regions back to their original spans. This is a strict operational check of diff understanding that goes beyond surface copying.
    \item[Z.] \textit{Diff Generation Task} --- \textit{diff} is unknown. This measures reliable diff synthesis: correct formatting and minimal, parseable edits. The outcome is directly relevant to code agents and patch-based evaluations where the system must emit a patch that tools can apply and reviewers can inspect.
\end{enumerate}

First, we apply \dsreal to analyze how well proprietary and open-source LLMs work with the most widespread unified diff format. 
Proprietary models consistently outperform open-source models, with Claude 4 Sonnet and GPT-4.1 achieving the highest scores across the tasks; GPT-4.1 is more sensitive to a task prompt and tends to emit V4A diff format by default. 
For open models, scores improve as model size grows, yet performance is still short of strong results, especially for smaller open models. \dsreal can help by providing focused, reproducible checks that highlight where formats and prompts make the most difference.
This analysis clarifies model behavior on the benchmark tasks and establishes strong reference results. 
Building on these findings, we then compare alternative edit representations: search-replace format consistently performs best across all tasks for stronger models, while udiff and udiff-h remain competitive alternatives. For smaller open models, performance remains limited regardless of format.

\section{Data Collection}

We construct \textit{\dsreal}, a compact benchmark of 1,000 real-world code edits derived from the CommitPackFT dataset \citep{commit-pack}, a large-scale corpus of open-source commits with paired before/after code and commit metadata.
Each instance in \dsreal is a triple $\langle \textit{old code}, \textit{new code}, \textit{diff} \rangle$, where \textit{diff} is computed between the two code versions.
This dataset serves as the common basis for all tasks presented in the next section.

To ensure diversity and quality, we apply systematic filtering and sampling.
We retain only commits that change a single file and exclude binary files, generated code, vendor directories, and trivial whitespace-only changes.
Changed files must contain at least 40 lines and no more than 1,000 lines in at least one version of the code (old or new). To balance edit complexity, we stratify commits by the number of changed hunks and change size (added + removed lines), targeting a 50/50 split between single-hunk and multi-hunk edits and a 40/40/20 distribution of small, medium, and large edits, defined by the 40th and 80th percentiles of change size (added + removed lines) within each bucket. In practice, edits with at most 7 changed lines are considered small (40th percentile), those with 8-24 lines medium (40th-80th percentile), and those with more than 24 lines large.
We also cap the number of examples per repository at 5 to encourage repository diversity.

The final dataset covers five programming languages, with 200 examples each: Python, JavaScript, and Java, which are widely used and appear in most existing benchmarks, and Kotlin and Rust, which are comparatively underrepresented but actively used in distinct contexts.

Most edits include both additions and deletions (81.5\%), while a minority are add-only (16.3\%) or delete-only (2.2\%).
In total, \dsreal spans 891 unique repositories.

\section{Tasks Description}

\subsection{Apply Task and Anti-Apply Task}
\textit{Apply} is a straightforward task: given the old code and the diff, generate the new code.
It can be viewed as a special case of code generation task with all the needed information being already in the prompt, yet the model has to correctly interpret the diff to solve it.
If an LLM is capable of understanding the diff and copying, we expect it to solve the task accurately, with larger models solving it near-perfectly.
In case a model scores low on the Apply task, it may require additional fine-tuning to work with diffs or an adaptation of the diff format.
Note that Apply is a natural simplification of the recently introduced problem of smart paste or fuzzy diff application with LLMs \citep{smart-paste}, \ie generate a new code state after applying a patch in some insufficient format.

\textit{Anti-Apply} is the problem complementary to the Apply task: given the new code and the diff, generate the old code.
Similar to the Apply task, all the data required to solve it is in the prompt.
In case there is a significant difference between the Apply and Anti-Apply scores for a model, it is a sign of model's overfitting on one of the tasks.

We evaluate performance on these tasks with slightly modified standard metrics. To calculate the following metrics for code snippets, we remove all the lines that contain only whitespace characters.

\begin{itemize}
    \item \textit{Stripped Exact Match} --- \textbf{EM} --- is 1 when two processed code snippets are exactly the same, 0 otherwise.
    \item \textit{Stripped Intersection over Union for lines} --- \textbf{IoU} --- ratio of unique lines in intersection to unique lines in union for two processed code snippets.
\end{itemize}

\subsection{Diff Generation Task}
\textit{Diff Generation Task} is formulated as follows: given the old code and the new code, generate the diff in the specified format.
This task is different from the first two, as it does not require generating long code sequences but rather shorter strictly formatted diffs.

The choice of format is crucial for the Diff Generation Task, as the model may be more or less aware of the given format. The unified diff format is the most common format used by GitHub and SWE-bench, and LLMs have higher chances of seeing it during pretraining. This may be not the case for other formats, so we always add a format description in the system prompt.

Unlike Apply and Anti-Apply, diff calculation is not uniquely defined. Even for the same code change, different but valid diffs can be produced, depending on factors such as the number of context lines. Because of that, we cannot apply similarity metrics like EM or IoU to the Diff Generation task. To overcome this problem, we evaluate EM and IoU for the \textit{new code} and the code that is the result of applying the generated diff to the \textit{old code}. Because diff application is a strict, all-or-nothing procedure, EM and IoU after application only capture exact correctness when the generated diff applies successfully. 
To measure partial correctness even when application fails, we also compute F1-scores over the sets of added and deleted lines \citep{info-retrieval}.

Finally, the standard unified diff hunk header contains line numbers, which are used to resolve uncertainties. However, it can be challenging to generate a hunk header before the hunk itself, and such uncertainties represent less than 1\% of the dataset. For this reason, we ignore hunk headers during patch application when they are not needed.

Here is the formal definition for metrics that we use to evaluate the Diff Generation task.

\begin{itemize}
    \item \textit{Parsing Rate} and \textit{Applying Rate} --- the fraction of generated diffs that can be parsed, and the fraction that can be successfully applied to the old code.
    \item \textit{EM} and line-level \textit{IoU} after application ––– EM and IoU values of the new code and the code after applying diff to the old code, 0 if the generated diff cannot be parsed.
    \item \textit{F1-score on Addition Lines} --- \textbf{F1+} --- F1-score between the set of added lines in the generated diff and those in the reference. 
    \item \textit{F1-score on Deletion Lines} --- \textbf{F1--} --- F1-score between the set of deleted lines in the generated diff and those in the reference.
\end{itemize}

\section{Unified Diff Evaluation}

In this section, we provide an extensive evaluation for the unified diff format on \dsreal. The results are organized in two subsections: proprietary models and open-source models.

All models are evaluated with a fixed user prompt for each task (see Appendix \ref{app:task-prompts}). We vary the system prompt in two ways:
\begin{enumerate}
    \item \textbf{w/o format} ––– a generic system prompt ("You are a helpful assistant.") with no explicit description of the edit representation.
    \item \textbf{w/ format} --- a system prompt that explicitly defines the unified diff format (see Figure \ref{fig:desc-system-prompt} in Appendix \ref{app:system-prompts})
\end{enumerate}

\subsection{Proprietary Models}
Proprietary LLMs are usually more powerful than open-source ones \citep{claude, gemini, gpt-4o, gpt4.1}, and since the tasks are conceptually simple, we expect the best models to achieve near-perfect scores.

Table \ref{tab:api-apply-antiapply-real} shows the results for the Apply and Anti-Apply tasks on the \dsreal dataset. Claude 4 Sonnet outperforms all models, making almost no mistakes on Apply under either system prompt, though it shows some quality drop on Anti-Apply. GPT-4.1 also performs very well across both tasks, but its results vary more across prompts.
For some models (\eg GPT-4.1), quality drops on Apply when moving from the \textit{w\slash o~format} to the \textit{w\slash~format} system prompt, while the scores on Anti-Apply remain unaffected. Such inconsistencies may indicate overfitting: the model focuses on the diff description and outputs a diff instead of code (see Figure \ref{fig:gpt4.1-diff-for-apply} in Appendix \ref{app:error-analysis}). 

Although diff application is a conceptually simple task, only the strongest proprietary models solve it perfectly. Open-source models, and even some smaller proprietary ones, still make mistakes. This confirms that the benchmark is well-calibrated: it exposes meaningful differences in models’ ability to interpret diff syntax, align edits with code, and produce the correct target version exactly.

\begin{table}[t]
    \centering
    \caption{Results of closed models on the Apply and Anti-Apply tasks on \dsreal.}
    \begin{tabular}{l l c  c  c  c}
        \toprule
        \multirow{2}{*}{\textbf{Model}} & \multirow{2}{*}{\textbf{Prompt}} & \multicolumn{2}{c}{\textbf{Apply}} & \multicolumn{2}{c}{\textbf{Anti-Apply}}\\
        \cmidrule(lr){3-4}\cmidrule(lr){5-6}
         & & \textbf{EM} & \textbf{IoU} & \textbf{EM} & \textbf{IoU}\\
        \midrule
        GPT 4o & w\slash o format            & 0.83 & 0.97 & 0.88 & 0.97 \\
        GPT 4o & w\slash format              & 0.87 & 0.98 & 0.89 & 0.98 \\
        GPT 4o-mini  & w\slash o format      & 0.70 & 0.96 & 0.57 & 0.85 \\
        GPT 4o-mini & w\slash format         & 0.05 & 0.12 & 0.34 & 0.49 \\
        \midrule
        GPT 4.1 & w\slash o format           & 0.92 & 0.98 & 0.93 & 0.98 \\
        GPT 4.1 & w\slash format             & 0.81 & 0.86 & 0.95 & 0.98 \\
        GPT 4.1 mini & w\slash o format      & 0.90 & 0.99 & 0.86 & 0.97 \\
        GPT 4.1 mini & w\slash format        & 0.90 & 0.98 & 0.85 & 0.97 \\
        GPT 4.1 nano & w\slash o format      & 0.38 & 0.85 & 0.03 & 0.63 \\
        GPT 4.1 nano & w\slash format        & 0.29 & 0.58 & 0.00 & 0.08 \\
        \midrule
        Claude 4 Sonnet & w\slash o format   & 0.95 & 0.99 & 0.87 & 0.90 \\
        Claude 4 Sonnet & w\slash format     & 0.96 & 0.99 & 0.87 & 0.89 \\
        Gemini 2.5 Flash & w\slash o format  & 0.91 & 0.97 & 0.77 & 0.85 \\
        Gemini 2.5 Flash & w\slash format    & 0.93 & 0.98 & 0.81 & 0.85 \\
        \bottomrule
    \end{tabular}
    \label{tab:api-apply-antiapply-real}
\end{table}

Table \ref{tab:api-diff-gen-real} shows the results for the Diff Generation task. Here, the effect of the system prompt is much stronger: explicitly describing the format narrows the model's choices and reduces the likelihood of switching to alternative diff formats.
This effect is especially noticeable for GPT 4.1, which, without explicit instructions, frequently outputs diffs in V4A, the format it was trained to produce, rather than unified diff (see Figure \ref{fig:gpt4.1-err-diff-gen} in Appendix \ref{app:error-analysis}).
\begin{table}[t]
    \centering
    \caption{Results of closed models on the Diff Generation task on \dsreal.}
    \begin{tabular}{l l c  c  c  c  c c}
        \toprule
        \textbf{Model} & \textbf{Prompt} & \textbf{EM} & \textbf{IoU} & \textbf{f1+} & \textbf{f1--} & \textbf{Apply Rate} & \textbf{Parsing Rate} \\
        \midrule
        GPT 4o & w\slash o format                         & 0.37 & 0.52 & 0.74 & 0.71 & 0.53 & 0.83 \\
        GPT 4o & w\slash format                           & 0.38 & 0.56 & 0.86 & 0.80 & 0.58 & 0.97 \\
        \rowcolor{gray!20}GPT 4o-mini & w\slash o format                    & 0.18 & 0.38 & 0.70 & 0.60 & 0.40 & 0.93 \\
        \rowcolor{gray!20}GPT 4o-mini & w\slash format                      & 0.17 & 0.28 & 0.66 & 0.56 & 0.29 & 0.91 \\
        \midrule
        GPT 4.1 & w\slash o format                        & 0.34 & 0.36 & 0.42 & 0.41 & 0.36 & 0.43 \\
        GPT 4.1 & w\slash format                          & 0.76 & 0.78 & 0.96 & 0.96 & 0.79 & 0.99 \\
        \rowcolor{gray!20}GPT 4.1-mini & w\slash o format                   & 0.12 & 0.12 & 0.15 & 0.14 & 0.13 & 0.16 \\
        \rowcolor{gray!20}GPT 4.1-mini & w\slash format                     & 0.60 & 0.64 & 0.74 & 0.72 & 0.64 & 0.77 \\
        GPT 4.1-nano & w\slash o format                   & 0.23 & 0.31 & 0.38 & 0.37 & 0.32 & 0.45 \\
        GPT 4.1-nano & w\slash format                     & 0.21 & 0.27 & 0.41 & 0.38 & 0.28 & 0.54 \\
        \midrule
        \rowcolor{gray!20}Claude 4 Sonnet & w\slash o format                & 0.62 & 0.65 & 0.76 & 0.75 & 0.65 & 0.79 \\
        \rowcolor{gray!20}Claude 4 Sonnet & w\slash format                  & 0.85 & 0.89 & 0.96 & 0.95 & 0.89 & 1.00 \\
        Gemini 2.5 Flash & w\slash o format               & 0.75 & 0.86 & 0.93 & 0.92 & 0.87 & 0.99 \\
        Gemini 2.5 Flash & w\slash format                 & 0.71 & 0.82 & 0.95 & 0.92 & 0.84 & 1.00 \\
        \bottomrule
    \end{tabular}
    \label{tab:api-diff-gen-real}
\end{table}

We observe that the Apply Rate is nearly identical to the IoU across models, indicating that whenever a model produces an applicable diff, it is usually very close to the expected result.

\subsection{Open Source Models}
To analyze how model size affects diff understanding, we evaluate the Qwen2.5-Coder series \citep{qwencoder}, a family of open-source, code-focused LLMs ranging from 0.5B to 32B parameters. Qwen2.5-Coder models achieve near state-of-the-art results on standard code generation and reasoning benchmarks among open-source models, making them a strong testbed for scaling analysis on \dsreal. Tables~\ref{tab:qwen-apply-antiapply-real} and~\ref{tab:qwen-diff-gen-real} report results for the Apply/Anti-Apply and Diff Generation tasks, respectively.

\begin{table}[t]
    \centering
    \caption{Results of the Qwen Coder family of models on the Apply and Anti-Apply tasks on \dsreal with \textit{w\slash format} system prompt.}
    \begin{tabular}{l c  c  c  c}
        \toprule
        \multirow{2}{*}{\textbf{Model}} & \multicolumn{2}{c}{\textbf{Apply}} & \multicolumn{2}{c}{\textbf{Anti-Apply}}\\
        \cmidrule(lr){2-3}\cmidrule(lr){4-5}
         & \textbf{EM} & \textbf{IoU} & \textbf{EM} & \textbf{IoU}\\
        \midrule
        Qwen2.5-Coder-0.5B-Instruct  & 0.00 & 0.39 & 0.00 & 0.46 \\
        Qwen2.5-Coder-1.5B-Instruct  & 0.11 & 0.39 & 0.05 & 0.70 \\
        Qwen2.5-Coder-3B-Instruct    & 0.36 & 0.81 & 0.15 & 0.71 \\
        Qwen2.5-Coder-7B-Instruct    & 0.59 & 0.94 & 0.64 & 0.93 \\
        Qwen2.5-Coder-14B-Instruct   & 0.82 & 0.97 & 0.82 & 0.97 \\
        Qwen2.5-Coder-32B-Instruct   & 0.85 & 0.98 & 0.86 & 0.98 \\
        \bottomrule
    \end{tabular}
    \label{tab:qwen-apply-antiapply-real}
\end{table}

\begin{table}[t]
    \centering
    \caption{Results of the Qwen Coder family of models on the Diff Generation task on \dsreal with \textit{w\slash format} system prompt.}
    \begin{tabular}{l c  c  c  c  c c}
        \toprule
        \textbf{Model} & \textbf{EM} & \textbf{IoU} & \textbf{f1+} & \textbf{f1--} & \textbf{Apply Rate} & \textbf{Parsing Rate} \\
        \midrule
        Qwen2.5-Coder-0.5B-Instruct & 0.00 & 0.00 & 0.00 & 0.02 & 0.01 & 0.23 \\
        Qwen2.5-Coder-1.5B-Instruct & 0.01 & 0.04 & 0.16 & 0.18 & 0.04 & 0.70 \\
        Qwen2.5-Coder-3B-Instruct & 0.00 & 0.04 & 0.26 & 0.22 & 0.06 & 0.84 \\
        Qwen2.5-Coder-7B-Instruct & 0.03 & 0.17 & 0.32 & 0.27 & 0.19 & 0.70 \\
        Qwen2.5-Coder-14B-Instruct & 0.14 & 0.35 & 0.63 & 0.53 & 0.38 & 0.92 \\
        Qwen2.5-Coder-32B-Instruct & 0.23 & 0.46 & 0.77 & 0.65 & 0.49 & 0.99 \\
        \bottomrule
    \end{tabular}
    \label{tab:qwen-diff-gen-real}
\end{table}

We observe a clear scaling trend: performance improves steadily with model size. On Apply and Anti-Apply, reliable results emerge around the 7B scale, with the largest Qwen Coder approaching GPT-4o. In contrast, none of the open-source models achieve comparable performance on Diff Generation. This gap suggests that handling diff syntax and formatting requires substantially more capacity than simply applying edits. It may also help explain why smaller models perform poorly on complex downstream benchmarks such as SWE-bench, where correctly generating patches is critical.

\section{Diff Format Exploration}
While the unified diff format is the most widely used, it is not the only way to represent edits. 
Different formats impose different structural constraints and tokenization patterns, which can affect how easily models generate, parse, and apply them. 
To probe these effects, we compare several representations side by side on \dsreal.
For each format, the system prompt included both a description of the format and one example (see Appendix~\ref{app:diff-formats-examples}).

We evaluate the following formats:
\begin{enumerate}
    \item \emph{udiff}: the standard unified diff (Figure~\ref{fig:udiff-example}).
    \item \emph{udiff-h}: unified diff with a relaxed hunk header, written as \texttt{@@~...~@@} (Figure~\ref{fig:udiff-h-example}).
    \item \emph{udiff-l}: unified diff with verbose line markers \texttt{ADD}, \texttt{DEL}, \texttt{CON} instead of the single-character \texttt{+}, \texttt{-}, and a leading space (Figure~\ref{fig:udiff-l-example}).
    \item \emph{search-replace}: a sequence of edits where each \emph{search} substring is replaced by a \emph{replace} substring, following \citet{Aider-udiff} and \citet{patchpilot} (Figure~\ref{fig:sr-example}).
\end{enumerate}

We include \emph{udiff} and \emph{search-replace} because they are widely used. The two \emph{udiff} variants address practical generation issues: \emph{udiff-h} avoids committing to exact line numbers before the hunk body is produced, and \emph{udiff-l} reduces ambiguity and token collisions by replacing single-character markers with explicit tags.

\begin{table}[!htp]
    \centering
    \caption{Results on \dsreal for various diff formats.
    For each model, the best-performing format is highlighted per task:
    \underline{Apply}, \textit{Anti-Apply}, and \textbf{Diff Generation}.
    If a format is best in multiple tasks, the corresponding styles are combined
    (e.g., \textbf{\underline{udiff}}).}
    \begin{tabular}{l l c  c  c  c  c}
        \toprule
        \multirow{2}{*}{\textbf{Model}} & \multirow{2}{*}{\textbf{Diff Format}} & \textbf{Apply} & \textbf{Anti-Apply}
        & \multicolumn{3}{c}{\textbf{Diff Generation}}\\
        \cmidrule(lr){3-3}\cmidrule(lr){4-4}\cmidrule(lr){5-7}
         & & \textbf{EM} & \textbf{EM} & \textbf{EM} & \textbf{f+} & \textbf{f--}\\
\midrule
        \multirow{4}{*}{ GPT 4o }
            & udiff                          & 0.86 & 0.85 & 0.41 & 0.90 & 0.84 \\
            & udiff-h                        & 0.85 & 0.89 & 0.30 & 0.80 & 0.72 \\
            & udiff-l                        & 0.82 & 0.88 & 0.01 & 0.02 & 0.02 \\
            & \underline{\textit{\textbf{search-replace}}} & 0.94 & 0.94 & 0.73 & 0.91 & 0.83 \\
\midrule
        \multirow{4}{*}{ GPT 4.1 }
            & udiff                          & 0.90 & 0.88 & 0.81 & 0.97 & 0.97 \\
            & udiff-h                        & 0.92 & 0.93 & 0.69 & 0.80 & 0.79 \\
            & udiff-l                        & 0.91 & 0.88 & 0.08 & 0.11 & 0.10 \\
            & \underline{\textit{\textbf{search-replace}}} & 0.96 & 0.93 & 0.95 & 0.97 & 0.93 \\
\midrule
        \multirow{4}{*}{ GPT 4.1-mini }
            & udiff                          & 0.89 & 0.81 & 0.78 & 0.90 & 0.88 \\
            & \textit{udiff-h}               & 0.86 & 0.85 & 0.69 & 0.84 & 0.85 \\
            & udiff-l                        & 0.86 & 0.83 & 0.01 & 0.02 & 0.02 \\
            & \underline{\textbf{search-replace}} & 0.94 & 0.84 & 0.92 & 0.95 & 0.93 \\
\midrule
        \multirow{4}{*}{ GPT 4.1-nano }
            & \textbf{udiff}                 & 0.37 & 0.02 & 0.50 & 0.82 & 0.78 \\
            & udiff-h                        & 0.32 & 0.01 & 0.44 & 0.79 & 0.75 \\
            & \textit{udiff-l}               & 0.44 & 0.04 & 0.00 & 0.00 & 0.00 \\
            & \underline{search-replace}     & 0.54 & 0.01 & 0.07 & 0.08 & 0.08 \\
\midrule
        \multirow{4}{*}{ Claude 4 Sonnet }
            & udiff                          & 0.95 & 0.82 & 0.82 & 0.97 & 0.95 \\
            & udiff-h                        & 0.95 & 0.87 & 0.83 & 0.97 & 0.95 \\
            & \textit{udiff-l}               & 0.95 & 0.93 & 0.66 & 0.83 & 0.82 \\
            & \underline{\textbf{search-replace}} & 0.97 & 0.87 & 0.94 & 0.95 & 0.90 \\
\midrule
        \multirow{4}{*}{ Gemini 2.5 Flash }
            & udiff                          & 0.90 & 0.34 & 0.72 & 0.94 & 0.94 \\
            & udiff-h                        & 0.91 & 0.54 & 0.76 & 0.95 & 0.94 \\
            & udiff-l                        & 0.79 & 0.07 & 0.36 & 0.47 & 0.46 \\
            & \underline{\textit{\textbf{search-replace}}} & 0.95 & 0.85 & 0.88 & 0.94 & 0.87 \\
\midrule
        \multirow{4}{*}{ Qwen2.5-Coder-0.5B }
            & \underline{\textit{\textbf{udiff}}} & 0.01 & 0.01 & 0.00 & 0.04 & 0.02 \\
            & udiff-h                        & 0.01 & 0.01 & 0.00 & 0.03 & 0.02 \\
            & udiff-l                        & 0.00 & 0.01 & 0.00 & 0.00 & 0.00 \\
            & search-replace                 & 0.00 & 0.00 & 0.00 & 0.00 & 0.00 \\
\midrule
        \multirow{4}{*}{ Qwen2.5-Coder-1.5B }
            & udiff                          & 0.16 & 0.04 & 0.00 & 0.08 & 0.09 \\
            & udiff-h                        & 0.16 & 0.05 & 0.01 & 0.09 & 0.11 \\
            & udiff-l                        & 0.11 & 0.03 & 0.00 & 0.00 & 0.00 \\
            & \underline{\textit{\textbf{search-replace}}} & 0.22 & 0.08 & 0.20 & 0.36 & 0.37 \\
\midrule
        \multirow{4}{*}{ Qwen2.5-Coder-3B }
            & udiff                          & 0.38 & 0.17 & 0.01 & 0.18 & 0.21 \\
            & udiff-h                        & 0.41 & 0.20 & 0.02 & 0.24 & 0.21 \\
            & udiff-l                        & 0.34 & 0.13 & 0.00 & 0.02 & 0.04 \\
            & \underline{\textit{\textbf{search-replace}}} & 0.41 & 0.29 & 0.14 & 0.27 & 0.23 \\
\midrule
        \multirow{4}{*}{ Qwen2.5-Coder-7B }
            & udiff                          & 0.58 & 0.65 & 0.06 & 0.42 & 0.34 \\
            & udiff-h                        & 0.57 & 0.64 & 0.05 & 0.32 & 0.26 \\
            & udiff-l                        & 0.57 & 0.62 & 0.00 & 0.00 & 0.00 \\
            & \underline{\textit{\textbf{search-replace}}} & 0.75 & 0.71 & 0.28 & 0.61 & 0.55 \\
\midrule
        \multirow{4}{*}{ Qwen2.5-Coder-32B }
            & \textit{udiff}                 & 0.84 & 0.87 & 0.23 & 0.74 & 0.64 \\
            & udiff-h                        & 0.83 & 0.86 & 0.17 & 0.50 & 0.44 \\
            & udiff-l                        & 0.83 & 0.82 & 0.01 & 0.02 & 0.02 \\
            & \underline{\textbf{search-replace}} & 0.92 & 0.85 & 0.68 & 0.91 & 0.82 \\
        \bottomrule
    \end{tabular}
    \label{tab:diff-formats}
\end{table}

The results in Table~\ref{tab:diff-formats} show that \emph{search-replace} is the most effective representation overall, especially for larger models, achieving strong results across all three tasks. Structured \emph{udiff} and \emph{udiff-h} formats remain reliable but are consistently outperformed by \emph{search-replace}. In contrast, \emph{udiff-l} performs poorly across most settings, suggesting that verbose line markers make the format harder for model rather than easier.
Overall, \emph{search-replace} offers a good balance between generation simplicity and faithful application for high-capacity models, while smaller open models still struggle regardless of representation.

We also observe that the most effective representation depends on both model scale and task step (generation vs. application). We hypothesize three interacting causes:
\begin{enumerate}
    \item \textbf{Local vs.\ global constraints.} \emph{search-replace} avoids format-level global constraints such as predicting line numbers in hunk headers (e.g., \texttt{@@ -a,b +c,d @@}), matching hunk lengths, and preserving the number of context lines. Each replacement stands on its own, so an error in one edit does not invalidate the rest. Larger models are better at locating distinctive anchors and ordering small, local edits, which increases parse/apply success.
    \item \textbf{Marker collisions.} Small models may confuse single-character udiff markers (\texttt{+}, \texttt{-}, leading space) with ordinary code characters. Replacing them with explicit tags (\texttt{ADD}/\texttt{DEL}/\texttt{CON}) in \emph{udiff-l} makes the control tokens unambiguous and rare; the model mainly needs to decide, for each line, which tag to use and then emit the line, an easier decision for weaker models. Nevertheless, our results suggest that this verbosity increases complexity rather than actually helping the models.
    \item \textbf{Header scaffolding \& distribution shift.} Standard udiff hunk headers (\texttt{@@ -a,b +c,d @@}) provide numeric anchors and implicit ordering cues. Even though our application step does not rely on these numbers, their presence appears to help models structure the patch: they act as scaffolding that encourages hunks to be segmented and emitted in file order. In \emph{udiff-h}, this scaffold is removed, which not only introduces a distribution shift relative to pretraining corpora (most public diffs include numbers) but may also increase the frequency of hunks being emitted out of order (e.g., an edit to line~120 preceding one to line~90). Both effects make the resulting patches less reliable, even though the format change is minimal.
\end{enumerate}

\section{Limitations and Future Work}

The tasks in \dsreal are simplified proxies rather than full downstream applications. Establishing a quantitative link between performance on these tasks and outcomes in real systems such as commit message generation, automated bug fixing, or code review is left for future work. As a result, our findings should not be interpreted as direct predictions of production performance.

All experiments use non-reasoning LLMs under greedy decoding. Evaluation of tool use, multi-step reasoning, sampling strategies, or best-of-\emph{n} decoding, is an important question that is out of scope of this work. The reported numbers therefore reflect single-pass behavior and likely represent conservative estimates rather than achievable upper bounds.

The set of edit representations we study is limited to the most popular ones. Some alternatives remain unexplored, including richer tree- or AST-based patches, structured search-replace variants, and error-tolerant or partially specified formats. Note, that they can be integrated into \dsreal without modifying the benchmark.

\section{Related Work}

\paragraph{Code generation benchmarks.}

A long line of benchmarks evaluate LLMs on code generation from natural language. \citet{humaneval} and \citet{mbpp} introduced HumanEval and MBPP, which remain the most widely used, with extensions such as HumanEval+ and MBPP+ \citep{evalplus}, HumanEval-XL \citep{humaneval-xl}, and multilingual variations such as MultiPL-E \citep{multiple}. BigCodeBench \citep{bigcodebench} emphasizes more complex and library-heavy tasks.
These benchmarks primarily test function-level synthesis and correctness, but do not address how edits are represented or applied.

\paragraph{Editing and issue-resolution benchmarks.}

More recent work has evaluated language models in editing settings. \citet{canitedit} introduces instruction-following edits, and \citet{codeeditorbench} evaluate debugging, polishing and translation tasks. At a larger scale, \citet{swebench} proposed SWE-bench, which tasks models with resolving real GitHub issues by producing patches that apply and pass tests. These benchmarks reflect realistic workflows, but include many factors, from retrieval and long context reasoning to semantic correctness and patch formatting.
These benchmarks capture realistic workflows, including patch application, but their evaluation mixes retrieval, long-context reasoning, semantic correctness, and patch formatting --- making it difficult to isolate the role of edit representation.

\paragraph{Positioning.}

Despite this breadth, existing benchmarks all assume a fixed diff format, typically unified diff. In reality, there are multiple representations in use: normal diff, context diff, unified diff, OpenAI's V4A diff format \citep{gpt4.1}, and search/replace schemes, including anchored replacements used by some agents.
Format is therefore a real variable, not a constant. Aider's benchmarks \citep{aider-polyglot} are notable for reporting differences between formats (whole file rewrite, search/replace, unified diff), but their evaluation was motivated by tool design choices rather than a systematic study of edit representations.

Our benchmark takes a complementary approach: it decomposes the end-to-end code editing pipeline problem and focuses specifically on edit representation in isolation. By holding task context fixed, it enables us to measure how models handle alternative formats under controlled conditions. This design also makes evaluation lightweight and cheap: unlike agentic frameworks such as SWE-bench, it requires no repository setup or execution harness, yet still targets a core capability that directly impacts downstream systems.

\section{Conclusion}

We introduced \dsreal, a benchmark for evaluating LLMs’ ability to handle code diffs across multiple formats. The tasks isolate core subproblems that arise in downstream systems while remaining simple enough to serve as controlled probes for more complex workflows.

Our main contribution is a set of focused tasks with clearly specified inputs and targets, paired with automatic metrics. This framework offers a reproducible setting for studying model behavior in diff-centric workflows. Possible extensions include connecting the benchmark to downstream tasks such as commit message generation, exploring corrupted or partial diff application, and incorporating structured code editing.

We also establish baselines for the unified diff format by measuring how LLMs process udiff inputs directly. Frontier models perform strongly on the provided tasks, suggesting the value of introducing more challenging instances within the same framework.

Beyond udiff, we compare several edit representations across a range of models. The resulting trade-offs can guide representation design. Future work will address the observed flaws and iterate toward formats that improve faithfulness and applicability.

\bibliographystyle{plainnat}
\bibliography{paper}

\appendix

\section{Inference Details}

\subsection{System prompts}
\label{app:system-prompts}
\begin{figure}[H]
    \centering
    \begin{tcolorbox}[colback=white, colframe=black, boxrule=0.5pt, width=0.5\linewidth, sharp corners]
        \centering
        \texttt{
        You are a helpful assistant.
        } \\
    \end{tcolorbox}
    \caption{\textit{w\slash o format} System Prompt}
    \label{fig:HA-system-prompt}
\end{figure}

\begin{figure}[H]
    \centering
    \begin{tcolorbox}[colback=white, colframe=black, boxrule=0.5pt, width=0.9\linewidth, sharp corners]
        \texttt{You are a helpful assistant.}\\

\texttt{When referred to unified diff format, the formatting must be as follows:}\\

\texttt{Do NOT include or start with Git headers like diff --git ... or index ..... Use POSIX unified diff with headers --- <old> and +++ <new>, hunks @@ -old\_start,old\_count +new\_start,new\_count @@. Prefix context with space, removals -, additions +. 1-based numbering, LF newlines, 1 context lines. New file: --- /dev/null. Deleted file: +++ /dev/null.} \\
    \end{tcolorbox}
    \caption{\textit{w\slash format} System Prompt}
    \label{fig:desc-system-prompt}
\end{figure}

\subsection{Task Prompts}
\label{app:task-prompts}

\begin{figure}[H]
    \begin{tcolorbox}[colback=white, colframe=black, boxrule=0.5pt, width=0.9\linewidth, sharp corners]
        \texttt{You need to write a code that is a result of applying the following diff in unified diff format to the following code snippet:}\\
        \\

        \texttt{Diff:}\\
        \texttt{\{diff\}}\\
        \\

        \texttt{Code:}\\
        \texttt{\{old\_code\}}\\
        \\

        \texttt{Use triple backtick formatting for you answer (e.g., \textasciigrave\textasciigrave\textasciigrave  python...\textasciigrave\textasciigrave\textasciigrave).}\\ 
    \end{tcolorbox}
    \caption{Prompt Template for the Apply Task}
    \label{fig:apply-prompt}
\end{figure}

\begin{figure}[H]
    \begin{tcolorbox}[colback=white, colframe=black, boxrule=0.5pt, width=0.9\linewidth, sharp corners]
        \texttt{You are given a code snippet that results from applying a unified diff. Your task is to reconstruct the original version of the code before the diff was applied.}\\
        \\

        \texttt{Diff:}\\
        \texttt{\{diff\}}\\
        \\

        \texttt{Code After Applying the Diff:}\\
        \texttt{\{new\_code\}}\\
        \\

        \texttt{Use triple backtick formatting for you answer (e.g., \textasciigrave\textasciigrave\textasciigrave  python...\textasciigrave\textasciigrave\textasciigrave).}\\ 
    \end{tcolorbox}
    \caption{Prompt Template for the Anti-Apply Task}
    \label{fig:anti-apply-prompt}
\end{figure}

\begin{figure}[H]
    \begin{tcolorbox}[colback=white, colframe=black, boxrule=0.5pt, width=0.9\linewidth, sharp corners]
        \texttt{You need to write a diff in unified diff format that transforms code snippet 1 to code snippet 2:}\\
        \\

        \texttt{Code Snippet 1:}\\
        \texttt{\{old\_code\}}\\
        \\

        \texttt{Code Snippet 2:}\\
        \texttt{\{new\_code\}}\\
        \\

        \texttt{Use triple backtick formatting for you answer (e.g., \textasciigrave\textasciigrave\textasciigrave  diff...\textasciigrave\textasciigrave\textasciigrave).}\\ 
    \end{tcolorbox}
    \caption{Prompt Template for the Diff Generation Task}
    \label{fig:diff-gen-prompt}
\end{figure}

\subsection{Examples of Diff Formats}
\label{app:diff-formats-examples}

\begin{figure}[H]
    \begin{tcolorbox}[colback=white, colframe=black, boxrule=0.5pt, width=0.9\linewidth, sharp corners]
        \begin{verbatim}
@@ -1,2 +1,4 @@
+import math
+
 def calculate_area(radius):
-    return 3.14159 * radius * radius
+    return math.pi * radius * radius
@@ -4,2 +6,0 @@
-def old_function():
-    return "deprecated"
        \end{verbatim}
    \end{tcolorbox}
    \caption{Example of a diff in \textit{udiff} format. This example is included in system prompt for diff exploration.}
    \label{fig:udiff-example}
\end{figure}

\begin{figure}[H]
    \begin{tcolorbox}[colback=white, colframe=black, boxrule=0.5pt, width=0.9\linewidth, sharp corners]
        \begin{verbatim}
@@ ... @@
+import math
+
 def calculate_area(radius):
-    return 3.14159 * radius * radius
+    return math.pi * radius * radius
@@ ... @@
-def old_function():
-    return "deprecated"
        \end{verbatim}
    \end{tcolorbox}
    \caption{Example of a diff in \textit{udiff-h} format. This example is included in system prompt for diff exploration.}
    \label{fig:udiff-h-example}
\end{figure}

\begin{figure}[H]
    \begin{tcolorbox}[colback=white, colframe=black, boxrule=0.5pt, width=0.9\linewidth, sharp corners]
        \begin{verbatim}
@@ -1,2 +1,4 @@
ADD import math
ADD
CON def calculate_area(radius):
DEL     return 3.14159 * radius * radius
ADD     return math.pi * radius * radius
@@ -4,2 +6,0 @@
DEL def old_function():
DEL     return "deprecated"
        \end{verbatim}
    \end{tcolorbox}
    \caption{Example of a diff in \textit{udiff-l} format. This example is included in system prompt for diff exploration.}
    \label{fig:udiff-l-example}
\end{figure}

\begin{figure}[H]
    \begin{tcolorbox}[colback=white, colframe=black, boxrule=0.5pt, width=0.9\linewidth, sharp corners]
        \begin{verbatim}
<<<<<<< SEARCH
    return 3.14159 * radius * radius
=======
    return math.pi * radius * radius
>>>>>>> REPLACE

<<<<<<< SEARCH

def old_function():
    return "deprecated"
=======
>>>>>>> REPLACE
        \end{verbatim}
    \end{tcolorbox}
    \caption{Example of a diff in \textit{search-replace} format. This example is included in system prompt for diff exploration.}
    \label{fig:sr-example}
\end{figure}

\section{Typical Errors}
\label{app:error-analysis}

\begin{figure}[H]
    \begin{tcolorbox}[colback=white,colframe=black,boxrule=0.5pt,width=0.9\linewidth,sharp corners]
        \begin{verbatim}
--- a/src/main.rs
+++ b/src/main.rs
@@ -20,5 +20,15 @@
 	let mut num = 0;
+	let mut prev = 0;
 
 	for c in roman_number.chars() {
-		num += roman_char_to_decimal(c);
+
+		let digit = roman_char_to_decimal(c);
+
+		num += digit;
+
+		if prev != 0 && digit > prev {
+			num -= 2 * prev;
+		}
+
+		prev = digit;
 	}
        \end{verbatim}
    \end{tcolorbox}

    \caption{
        Typical incorrect answer for GPT 4.1 with \textit{w\slash format} system prompt in \emph{Apply Task}.
        Diff instead of new code is generated for more than 10\% of data points.
        For these data points, answer with \textit{w\slash o format} system prompt is correct, 
        and answer for Anti-Apply is correct with either system prompt.
    }
    \label{fig:gpt4.1-diff-for-apply}
\end{figure}

\begin{figure}[H]
  \centering
  
    \begin{tcolorbox}[colback=white,colframe=black,boxrule=0.5pt,width=0.9\linewidth,sharp corners]

      \begin{verbatim}
--- snippet1.py	2024-06-13 12:00:00.000000000 +0000
+++ snippet2.py	2024-06-13 12:00:00.000000000 +0000
@@ setup(
-    classifiers=[
-        'Development Status :: 4 - Beta',
+    classifiers=[
+        'Development Status :: 5 - Production/Stable',
      \end{verbatim}
    \end{tcolorbox}
\vspace{0.6ex}        
    \begin{tcolorbox}[colback=white,colframe=black,boxrule=0.5pt,width=0.9\linewidth,sharp corners]

        \begin{verbatim}
--- old
+++ new
@@ -18,7 +18,7 @@
     packages=['drupdates', 'drupdates.tests'],
     classifiers=[
-        'Development Status :: 4 - Beta',
+        'Development Status :: 5 - Production/Stable',
         'Environment :: Console',
         'Intended Audience :: Developers',
         'Intended Audience :: System Administrators',
        \end{verbatim}
  \end{tcolorbox}

  \caption{
        \textbf{Top Box.} Typical incorrect answer for GPT 4.1 with \textit{w\slash o format} system prompt in \emph{Diff Generation Task}.
        \textbf{Bottom Box.} Answer with \textit{w\slash format} system prompt.
        Incorrect diff is in V4A format instead of the unified diff format. 
        Such answers are generated for more than 50\% of data points.
    }
  \label{fig:gpt4.1-err-diff-gen}
\end{figure}

\end{document}